# Gender Differences in Research Topic and Method Selection in Library and Information Science: Perspectives from Three Top Journals

Chengzhi Zhang, Siqi Wei, Yi Zhao, Liang Tian
Department of Information Management, Nanjing University of Science and Technology, Nanjing, 210094, China
Corresponding author: Chengzhi Zhang (zhangcz@njust.edu.cn).

**Abstract:** Research in the social sciences has shown that there are gender differences in the selection of research methods, with women often opting for qualitative methods while men prefer quantitative methods. However, it is important to consider that research methods are generally chosen based on the research topic. To figure out the influence of gender on research method selection, a study was conducted in the field of Library and Information Science, using a more fine-grained method classification system and an automatic classification model called CogFT, which is based on full-text cognition. The findings showed that women tend to use Interview while men prefer Theoretical approach, across a range of topics. The study offers insights into the specific research design processes that contribute to gender differences in method selection and suggests ways to promoting gender inclusivity and equality in academia by considering research method use and guidance.

## 1. Introduction

Gender disparities are evident in the scientific community, with women being underrepresented in terms of scientific output, academic rank, and influence (Holman et al., 2018; Larivière et al., 2013; Mairesse & Pezzoni, 2015). These disparities cannot be entirely attributed to physical differences or external prejudice against women (Justman & Méndez, 2018; Lindberg et al., 2010). One major factor contributing to this issue is the differing career preferences and choices of men and women. In academia, these differences manifest themselves in varying research field and topic choices (Ceci & Williams, 2011; Thelwall et al., 2019). For instance, fewer women than men choose to pursue further education in areas that need advanced math skills (Ceci et al., 2014). The gender difference in topic selection also varies across disciplines; for example, women in education prioritize primary and secondary schooling, whereas women in management emphasize social and human-centered research (Scharber et al., 2019; Wullum Nielsen & Börjeson, 2019). Several studies attempted to explain the differences in topic choice by the dimension of interest, i.e., women's interests are typically tied to people, whereas men's interests are usually related to things (Su et al., 2009; Woodcock et al., 2013). However, a single viewpoint cannot uniformly explain every field, as demonstrated by the underrepresentation of women in computer science sub-fields such as human-computer interaction and natural language processing, which both involve people (Su & Rounds, 2015).

Women and men might choose topics differently, which in turn influences how they choose research methodologies. Evidence suggests a connection between the gender of the author and the method of study (Diaz-Kope et al., 2019; Williams et al., 2018). Women prefer qualitative methods, whereas men typically utilize quantitative methods to problem-solving (Ashmos Plowman & Smith, 2011; Nunkoo et al., 2020). From a social epistemological perspective, the reason for the differences between female and male research paradigms is that women and men may perceive the world in different ways (Oakley, 2000; O'Shaughnessy & Krogman, 2012; Rolin, 2004), which challenges the research epistemology and the notion that research methods are gender neutral but illustrates that social dimensions, such as gender, influence the production of knowledge in the sciences (Diaz-Kope et al., 2019; Rolin, 2004; Williams et al., 2018). The present study explores gender differences in research topics and methods within a particular field to analyze and understand the potential reasons for gender imbalances in academia and assess the impact of gender on knowledge creation.

## 2. Problem statement

While the conventional view is that research methods are chosen based on the nature of research topic and objectives (Creswell, 1994, 2003), suggesting gender neutrality of method selection, research in certain social science fields has shown that gender may also play a role in methodological design (Diaz-Kope et al., 2019; Williams et al., 2018). The contradiction between theory and findings may be challenged by the fact that these studies do not adequately consider the influence of research topics on the method selection. Additionally, current research has focused on the differences in the choice of qualitative and quantitative methods between women and men; there is still a lack of statistical data on gender differences in the choice of specific research methods like questionnaires and experimental design.

To further support the earlier research findings and to gain a deeper understanding of gender differences in methodological choices from specific research design processes, gender differences in specific method usage within different research topics, using the field of library and information science (LIS) as an example, was investigated. It is noteworthy that women are still underrepresented in top LIS journals, particularly in those centered on information science topics (Lund & Shamsi, 2021). LIS was selected for its wide range of topics and methods, but this characteristic also posed a challenge when it came to evaluating the methodological categories of research articles. Traditional content analysis methods can be demanding for annotators and can consume considerable time. To address these issues, a full-text cognitive-based classification model of research methods called CogFT (a detailed description is provided in section 4.3) was been introduced. The model allows for the identification of relevant parts of research method descriptions in the full text and the extraction of specific research method categories. What sets CogFT apart is its ability to classify research methods with a small amount of annotated corpus, making it ideal for this difficult task.

Based on the above objectives and methods, the study examines the following three research questions:

***RQ1***: Are there differences in research topic and method selection between male and female scholars in the field of LIS?

***RQ2***: Are there differences in research method selection between male and female scholars within different LIS research topics?

***RQ3***: How do male and female scholars differ in their choice of research topics and methods in different LIS journals?

## 3. Literature review

### *3.1 Gender differences among scholars in specific fields*

Gender differences are prevalent in all fields and the manifestation of gender differences varies from area to area (Thelwall, 2020). In most STEM (Science Technology Engineering Mathematics) fields where mathematics is the core foundation, such as physics and computer science, men are significantly better represented than women, while in fields where mathematics is used as a secondary tool, such as psychology and social sciences, female practitioners tend to outnumber men (Ceci et al., 2014; Su & Rounds, 2015; Thelwall et al., 2019). But across the scientific spectrum, female academics lag behind their male counterparts in terms of both academic output and the attainment of senior titles (Larivière et al., 2013; Mairesse & Pezzoni, 2015). Although the number of female academics has increased after a long period of appeal and improvement, the gender gap in terms of academic status and academic influence has remained and is not likely to narrow anytime soon, especially in the fields of surgery, computer science, physics and mathematics (Hart et al., 2022; Holman et al., 2018; Mohammad, 2020).

The gender gap in the number of practitioners in LIS is smaller than it is in the aforementioned fields. Earlier studies have shown that the ratio of female to male academics in LIS is virtually equal, or that the female academic ratio is slightly greater than the male academic ratio (Herubel, 1992; Nisonger, 1996). However, there are also pertinent data that highlight a difference between male and female academics in terms of citation acquisition, propensity, and collaborative propensity (Håkanson, 2005; Reece-Evans, 2010). With the growth of information science, the gender gap in LIS academia has been widening (Bisaria & Jaiswal, 2019; Lund & Shamsi, 2021), and as LIS is an interdisciplinary area, this gap cannot be disregarded.

### *3.2 Differences in research topic and method selection*

Topic choice in academic research not only reflects the paper's content and aim but also reveals the author's research interests and direction. A highly cited study by Hoppe and colleagues (2019) showed that topic choice plays a role in the unequal distribution of funding for African-American and Black scientists, leading to controversy. This finding highlights disparities in topic choices among different groups and their potential impact on research development. Gender-based differences in topic choices have long been observed. Women in natural language processing study dialogue, discourse, and emotions, while men focus on formal semantics, analysis, and finite state models. Similarly, female scholars in LIS publish more articles on usage research while male scholars focus on bibliometric analysis (Bisaria & Jaiswal, 2019; Vogel & Jurafsky, 2012). Women tend to study topics related to people, while men prefer topics related to things, which may contribute to differences in citation rates (Su et al., 2009; Thelwall et al., 2019).

Choosing the right research method is crucial for ensuring the quality of academic research. While less attention is given to differences in research method usage due to the lack of a universally applicable classification system, methods are generally divided into qualitative or quantitative, or single or mixed methods (Creswell, 1994, 2003; Fidel, 2008; Oakley, 2000). Most studies focus on these two divisions, with some revealing that female scholars tend to prefer qualitative methods while male scholars favor quantitative ones (Ashmos Plowman & Smith, 2011; Nunkoo et al., 2020; Williams et al., 2018). A study on career age found that junior researchers were more likely to use mixed research methods, while their senior counterparts tended to employ theoretical approaches (Lou et al., 2021). Another study showed that the research method usage also varied between countries, for example, Belgian and Netherlands scholars showed a preference for bibliometric methods (Zhang & Tian, 2023).

### *3.3 Classification system construction and automatic classification of research methods*

Research methodology is a vital element in the development of a discipline, and the standardization of research methods indicates its maturity. Therefore, constructing a classification system for research methods holds significant importance for any field. In the 1990s, Järvelin & Vakkari (1990, 1993) conducted content analysis on top LIS journals to analyze research strategies and data collection methods, proposing a benchmark classification system consisting of 13 research methods. Chu & Ke (2017), manually inspected and annotated research papers from 2001 to 2010 in three LIS core journals, determining 16 research methods based on data collection techniques. Most studies on research methodology in the field of LIS have either utilized one of these two method systems, enhanced one, or created an optimized system by incorporating both systems (Lou et al., 2021; Lund & Wang, 2021; Ma & Lund, 2021).

Most studies use content analysis to identify and define research methods, but this is a laborious and time-consuming task that hampers analysis and mining of methods in larger publications. Some academics are developing machine learning and deep learning methods to automatically classify research methods based on existing classification systems and annotated datasets. Two primary types of corpora are used for this: one uses paper titles and abstracts, and the other uses full-text papers. Eckle-Kohler et al. (2013) classified research methods using a multi-label classification algorithm and abstracts of 1992 journal articles in the social sciences, achieving up to 0.68 F1 values for 15 methods. They suggested full-text content could improve automatic classification. Zhang et al. (2021b) tried a rule-based approach, noting the classification of research methods is a challenging task. Zhang & Zhang (2021) constructed a more hierarchical and detailed Chinese intelligence method classification system by semi-automated clustering of method entities in full-text papers.

## 4. Data and methodology

### *4.1 Data sources*

The data used in this study came from 5,281 research articles published in *Journal of the Association for Information Science and Technology* (*JASIST*), *Journal of Documentation* (*JDoc*) and *Library and Information Science Research* (*LISR*) from 1990 to 2019. These three journals were selected for their status as well-established, authoritative, international LIS core journals and their unique and distinct focus on LIS research. JASIST's research focuses on a range of processes, from the production of information to its evaluation, along with the relevant tools and techniques involved. JDoc emphasizes theoretical and philosophical research in information sciences, while LISR concentrates on the research process and the practical implications of research findings for LIS. Due to the distinct research foci of the three journals, the data encompasses publications covering a wide range of topics and utilizing various methods. Moreover, the particular focus of each journal also attracts a specific type of contributing researcher, further contributing to the examination of gender differences in topic and methodological choices. These journals have also been selected in previous related studies (Chu, 2015; Chu & Ke, 2017; Zhang & Tian, 2023), and the selection of these journals in this study can enrich the findings of existing studies.

Only research publications were included in the data; editorials, book reviews, and other non-research items were excluded. This was because nearly all research articles provide a methodology description, facilitating the categorization of research methods. Over the past few decades, the research topics and methods in the field of LIS have been repositioned (Järvelin & Vakkari, 2021; Ma & Lund, 2021). Considering variations in research topics and methods over time, journal articles spanning a period of 30 years, from the early 1990s to the late 2010s, were collected. The number of research articles in each journal is shown in Table 1. *JASIST* had the highest number of publications, followed by *JDoc*, while *LISR* had the lowest, which may correlate with each journal's annual publication volume.

Table 1. Number of articles in *JASIST*, *JDoc* and *LISR*.

| **Journal** | **#(Article)** |
|---|---|
| *JASIST* | 3537 |
| *JDoc* | 1102 |
| *LISR* | 642 |
| Total | 5281 |

*4.2 Identification of research topic*

A distributed topic vector model called Top2Vec was selected to identify the research topic of articles from *JASIST*, *JDoc*, and *LISR* journals. Compared to popular probabilistic topic models, Top2Vec generates more informative and representative topics from the trained corpus (Angelov, 2020). The full text of articles serves as the initial corpus for topic modeling. The number of topics was set to be 17, based on a context-vector-based coherence metric (CV) and considering the category number of research methods (Appendix A.1). For each topic, the Top2Vec model generates the top 10 descriptive keywords, which are ranked based on cosine similarity-based semantic similarity to the topic (Appendix A.2). The topics were then named by reviewing these keywords and combining them with the knowledge of the discipline of LIS (Table 2).

Table 2. Identified topic names.

| |
|---|
| Information seeking behavior |
| Information retrieval/Models and algorithms |
| Information ethics/Philosophy of information |
| Science evaluation |
| Text processing |
| School libraries/Information literacy |
| Bibliometrics/Scientometrics |
| Health information/Communities |
| Information services/ Information behavior |
| Government policy/statutes |
| Organizational information activities/Knowledge management |
| Open access publishing |
| Information organization/ Information management |
| Technology adoption/ Behavior issues |
| Science of LIS |
| Personal digital curation/ Personal information management |
| Webometrics |

*4.3 Classification of research methods*

Given the difficulty of manually discerning research methods used in articles, an automatic classification model of research methods was developed. The classification labels for the automatic classification task were based on Chu & Ke (2017)'s classification system for LIS research methods, which was constructed using content analysis. This system considers data collection and analysis as two fundamental attributes of research methods and names the methods based on their respective data collection techniques. The system comprises 16 methods, including bibliometrics, content analysis, delphi study, ethnography/field study, experiment, focus groups, historical method, interview, observation, questionnaire, research diary/journal, theoretical approach, think aloud protocol, transaction log analysis, webometrics, and other methods. For detailed definitions of these research methods, please refer to the Appendix B.1. To facilitate this task, Chu & Ke (2017) generously provided a training corpus of 1,981 research articles published between 2001 and 2010 in *JASIST*, *JDoc*, and *LISR* journals, each of which was manually annotated with the corresponding research method.

Since research methods may not be fully described in the abstracts of academic papers, the full text was used for automatic classification. However, the full text is usually lengthy and contains a lot of noise unrelated to the research methodology, so it is necessary to extract the key contents of the research methodology from the full text. Inspired by CogLTX (Cognize Long TeXts) (Ding et al., 2020), an automatic classification model of research methods based on full-text cognition named CogFT (Cognize Full Texts) was developed (Tian, 2023). The model can be divided into two stages: full-text summary and automatic classification of research methods. First, the article is segmented into blocks, and then the key blocks related to research methods are extracted. Subsequently, the model uses those key blocks to classify the methods (Fig. 1).

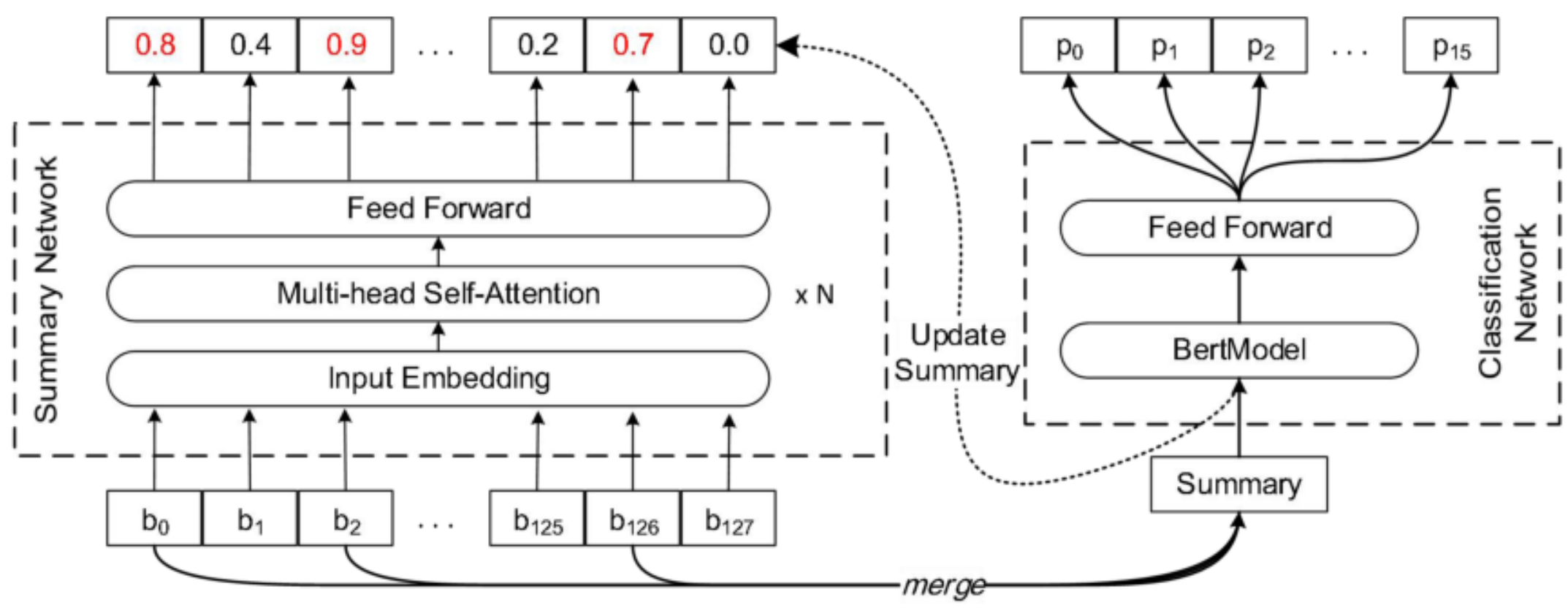

Fig. 1. Framework and implementation steps of the CogFT model(Tian, 2023).

In the full-text summary stage, the full text can be divided into 128 chunks of 64 length each. Therefore, the classification model can handle text up to 8192 in length. If the length of the text exceeds 8192, the text is intercepted from the front. SciBERT (Beltagy et al., 2019) is used for input embedding. The relationships between each block in the full text are then determined using a self-attention based global encoder. Finally, the summary network outputs the probability that each block is critical to the description of the research method. At the stage of classification of research methods, the four blocks with the highest probability and the top four blocks in the full text are input to the classification network, which uses SciBERT and Sigmoid layer for classification.

To evaluate the performance of the CogFT model in the automatic classification task of research methods, several additional models were chosen as baselines for comparison. These models include: 1) MS (Match selection) (Zhang et al., 2021b), 2) Ngram+Chi2+NB (CC) (Wang et al., 2021), 3) SciBERT-Truncation (C. Sun et al., 2019), 4) SciBERT-Hierarchical (Pappagari et al., 2019), and 5) BigBird (Zaheer et al., 2020). The classification of research methods involves a multi-label classification of texts, which can be evaluated using two types of metrics: sample-based metrics and label-based metrics (Zhang & Zhou, 2014). Both sample-based metrics (Samples_P, Samples_R, Samples_F1) and label-based metrics (Micro_P, Micro_R, Micro_F1, Macro_P, Macro_R, Macro_F1) were employed.

Prior experiments have shown that method categories with fewer than 300 articles performed poorly in automatic classification, such as Delphi study. To address this issue, the keywords of the low-frequency method categories were used to search the Web of Science for relevant papers to add to the training corpus, which were manually checked and annotated. These keywords were extracted from articles using TextRank (Mihalcea & Tarau, 2004). However, the process of data expansion inevitably introduced some mislabeled or noisy data. To mitigate this, confident learning was employed to identify and remove noisy data by estimating the joint distribution of noisy and true labels (Northcutt et al., 2021). Finally, the performance of classification models was evaluated using the larger dataset (Appendix B.2).

The superiority of the CogFT model in research method classification is evident, making it the ideal choice for categorizing the methods used in the articles collected for this study. The results obtained from the model can be found in Appendix B.3. It is significant to note that the total frequency of methods (6902) in the classification results exceeds the total number of articles collected (5281). This occurrence arises from the fact that certain articles encompass multiple research methods within their study design, indicating the utilization of mixed research methods. These articles constitute approximately 21.45% of the total dataset.

### *4.4 Gender inference of author*

In this paper, the authors' gender is inferred primarily by matching their names with publicly available gender datasets. Three different datasets were utilized for this purpose: 1) names and genders of newborn babies from 1880-2019 as published by the US Social Security Administration (*Popular Baby Names,* 2022); 2) a list of 9,300,182 PUBMED authors' names and their gender published by Torvik & Smalheiser on their agency's official website (Torvik & Smalheiser, 2009); and 3) World Gender Name Dictionary (WGND) by Raffo & Lax-Martinez at Harvard Dataverse (Raffo & Lax-Martinez, 2018). The datasets only include first names, and not surnames. Statistical filtering is required for the first two datasets, with a name being classified as closely associated with a particular gender only if more than 95% of individuals with that name correspond to that gender. For the WGND, the subset wgnd_noctry was used, as it has been purged of any conflicting gender-name information present in the WGND. To ensure the precision of gender matching using the three datasets, a hand-annotated list of authors' names and genders in the ACL Anthology by Vogel & Jurafsky(2012) was employed (Table 3).

Table 3. Precision of matching gender using datasets.

| **Dataset** | **#(Female name)** | **#(Male name)** | **Precision** |
|---|---|---|---|
| USSA | 59,743 | 33,509 | 0.9794 |
| PUBMED | 22,560 | 28,181 | 0.9772 |
| WGND | 98,646 | 67,119 | 0.9817 |

The three datasets were applied to match the gender of authors in LIS journal articles. To ensure accuracy, the gender inference relied on a consensus of two or more matches. However, the recall rate was only approximately 73%, meaning that some author genders were not identified through this process. For the remaining authors, a manual search was conducted to determine their gender. To locate author's profile photo, the personal pronouns used in the author's profile or in third-party reviews, a search was conducted on the search engines or academic websites like Bing, ResearchGate, and Google Scholar using the author's name, institution, email address, and other relevant information.

## 5. Results

This study focuses mainly on first authors, as they were deemed to play a key role in research topic and method selection (Lu et al., 2022). Only 281 papers (5.32%) had a discrepancy between the first and corresponding authors, indicating minimal impact of disagreements between the main contributor and the supervisor on the research's development and design. Of the 5,281 first authors of LIS journal papers examined, 5,273 (99.84%) were gender-identifiable, with 3,199 (60.67%) male and 2,074 (39.33%) female. The statistics are shaped by several factors, including the volume of articles published (Table 1) and the gender ratio of authors in different journals (Fig. 2).

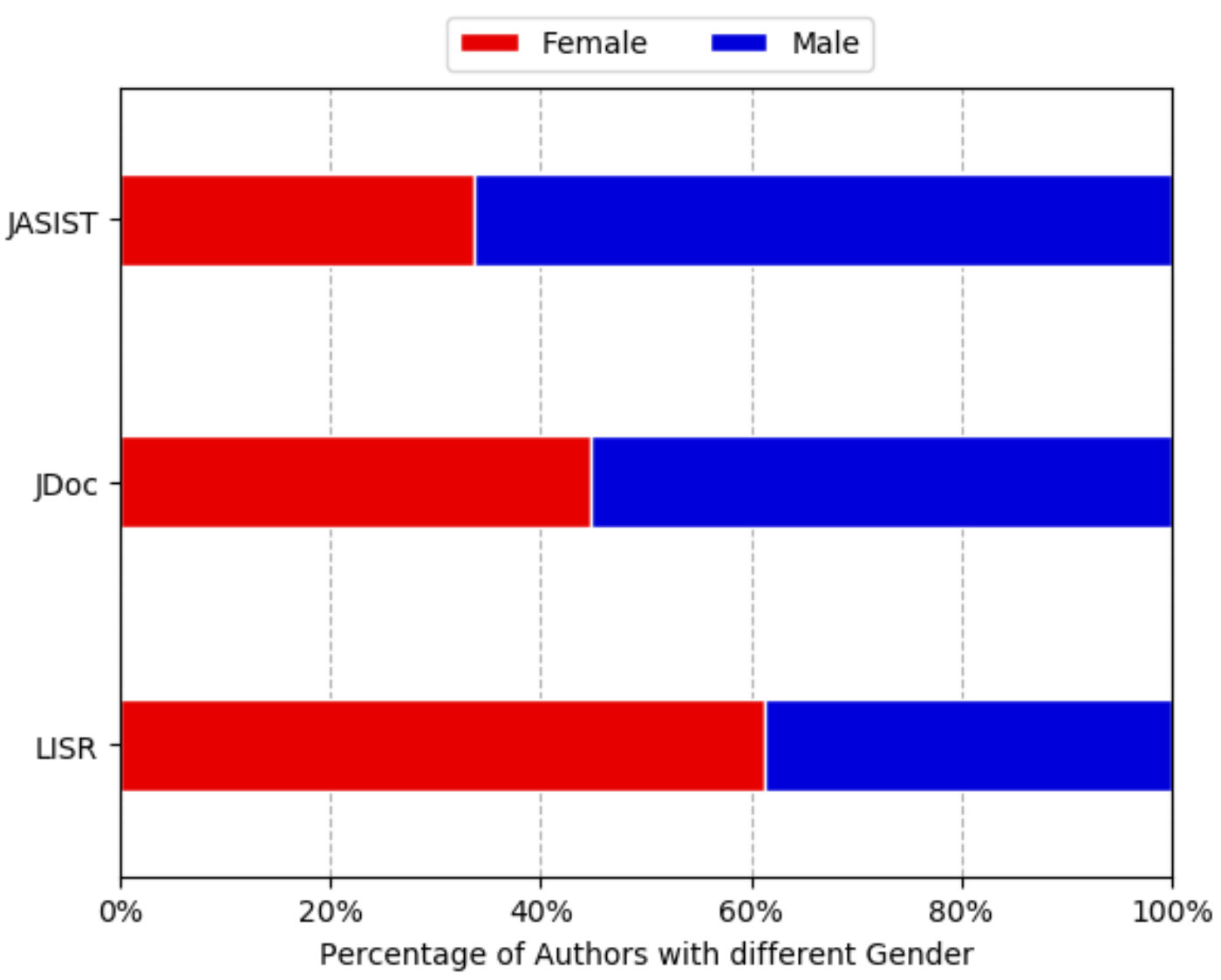

Fig. 2. Percentage of gender-specific authors for JASIST, JDoc and LISR.

Based on the gender inference findings, this section sequentially addresses the three RQs.

*5.1 Gender differences in LIS topic and method selection*

When compared to the overall gender share of authors (F/M≈4/6), the gender share of authors for 14 out of the 17 LIS research topics showed significant differences (Appendix C.1). Female authors had greater representation in seven topics, including information seeking behavior and school libraries/information literacy, while male authors were more prevalent in seven other topics, such as information retrieval/models and algorithms and information ethics/philosophy of information.

The chi-squared value indicates the degree of bias in favor of a topic selected by male or female authors in comparison to the overall distribution. After analyzing the chi-squared value rankings, health information/communities and school libraries/information literacy were the preferred research topics among female authors, whereas information retrieval/models and algorithms and science evaluation were the top choices for male authors.

The distribution of the 16 research methods used in LIS journal articles by male and female authors was counted in a similar manner (Appendix C.2). Note that an article can use multiple methods, thus the total frequency of methods used is not equal to the total number of articles. To satisfy the independence requirement of the chi-square statistic, the expected frequency used was the number of articles written by female and male authors, instead of the total frequency of all methods used by them. The chi-square value was then calculated independently for each method.

When compared to the total author gender ratio, 8 of the 16 research methods were substantially overrepresented by female authors at the 0.05 level of significance, while 3 methods were considerably overrepresented by male authors. Among the top five methods ranked by chi-square values, female authors tended to favor interview, questionnaire, and observation, while male authors showed a preference for bibliometrics and theoretical approach.

*5.2 Gender differences in LIS method selection within different topics*

In order to discover gender differences in scholars' choice of methods within the same topics, this study constructed a two-dimensional heat map based on the Percentage of Female Authors (PFA) in various topics and methods (Appendix C.3). The square in the graph is displayed in red if the PFA is larger than 0.5, and in blue if the PFA is less than 0.5. Frequencies under 5 were smoothed in the PFA calculation to prevent the appearance of extremely big or extremely small outliers. Meanwhile, all ratios were normalized to remove the impact of category imbalances (more men than women). Topics and methods were reordered depending on the chi-square value, with those with a higher PFA appearing in the top left corner of the graph, and those with a lower PFA appearing in the bottom right.

After normalization, 44.85% of squares were red (PFA>0.5) and 36.4% were blue (PFA<0.5). Median distinction showed significantly more dark squares (PFA>0.67) in red squares (19.67%) than dark squares (PFA<0.33) in blue squares (11.11%), suggesting female authors have a more pronounced bias in topic and method choices after adjusting for total headcount gap. Interview had the most red squares, and theoretical approach had the most blue squares. They also had the deepest red squares (PFA = 0.82) and darkest blue colors (PFA = 0.17), respectively. Speculation arises that, disregarding topic preference, female authors may have a stronger preference for interview, while male authors may favor theoretical approach. In some topics, such as health information/communities, information retrieval/models and algorithms, and science evaluation, which had a single color in their rows, method preferences of male or female authors were aligned with the absolute preferences for the topics. Further statistical analysis is needed to fully examine method preferences.

Six methods rejected the original hypothesis (p<0.05) that there is no significant difference between the proportion of male or female authors choosing a method in a given topic and the proportion of male or female authors in that topic, as determined by the chi-square statistics. Appendix C.4 displays the differences of the six methods used by female or male authors within topics. The circle size represents the chi-squared value, with red circles indicating methods that were used more by females and blue circles that were used more by males. The backgrounds are colored accordingly, with red for female-preferred topics, blue for male-preferred, and grey for no significant difference.

As observed in the vertical columns, female authors displayed a strong inclination towards using interviews across five different topics, even in the typically male-favored topic of information organization/information management. On the other hand, male authors displayed a greater tendency to utilize theoretical approaches in four specific topics that are dominated by females. Further, female authors exhibited a preference for four other research methods, including the questionnaire and observation, in one or two topics. Overall, the data suggests that female authors have a wider range of significantly preferred research methods compared to male authors.

*5.3 Gender differences in LIS topic and method selection in different journals*

*JASIST*, *JDoc*, and *LISR*'s diverse research focuses attract distinct contributors, causing a gender ratio disparity across these journals (Fig. 2) evident in the distribution of research topics and gender-specific authors within these topics (Appendix C.5). In *JASIST*, five primary topics accounted for more than half of all articles (51.94%), with male authors dominating four of these topics, except for information seeking behavior. *JDoc* had a considerable number of articles on information ethics/philosophy of information (20.78%), with more male authors in this topic. *LISR* published fewer articles, but female authors contributed more papers, with the top two topics (32.73%), school libraries/information literacy and health information/communities, having significantly more female authors than male authors at a 0.05 significance level. Notably, the high-frequency research topics in different journals hardly overlap.

There are also some differences in the use of research methods across the three journals. Appendix C.6 illustrates the distribution of research methods used by male and female authors in articles published in *JASIST*, *JDoc*, and *LISR*. Experiment was the most commonly employed method in *JASIST* and was utilized more often by male authors, likely due to the information retrieval/models and algorithms topic having the highest number of papers, which typically require this method. In *JDoc*, theoretical approach was the most frequently used method, with male authors employing it more often than their female counterparts, and usually required for the topic of information ethics/philosophy of information. In *LISR*, questionnaire, content analysis, and interview were all considerably more frequently used methods, and female authors tended to favor all three.

While the top two methods used differed between journals, experiment, theoretical approach, questionnaire, content analysis, interview, and bibliometrics were consistently among the top six methods utilized in each journal. Chi-square statistics were conducted on the frequency of use of these six methods in each journal, revealing that male authors had a notably higher tendency to utilize theoretical approach while female authors remarkably preferred interview, regardless of the journal ($p<0.01$). In *JASIST* and *JDoc*, two journals with a low percentage of female authors, female authors clearly favored the use of content analysis and questionnaire ($p<0.05$). Additionally, despite the stereotype that more male authors use experiment, the findings reject this idea, showing no discernible difference in the proportion of male authors using experiment and the overall proportion of male authors in each journal at a 0.05 level of significance.

## 6. Discussion

*6.1 Analysis of potential reasons for gender differences*

The gender differences in the research topic selection in the field of LIS may be shaped by the interdisciplinary nature of the discipline itself. According to Järvelin and Vakkari(2021), the focus of LIS research has evolved over time, with scholars increasingly collaborating with other fields to study major research topics. For example, the topic of information retrieval/models and algorithms is a product of collaboration with the field of computer science, and the topic of health information/communities draws from the field of medicine or health sciences. This could imply that the gender ratio in these classic LIS topics is influenced by the high participation of women in medicine or health Sciences as well as the over-representation of men in computer science, leading to a widening of the gender gap. However, this merely explains the extrinsic causes of gender disparities in LIS, not their true internal roots.

The traditional "person versus thing" theory, which holds that women are more interested in topics related to people and men explore more topics related to things (Thelwall et al., 2019), has been widely used in studies to explain gender differences in topics from a social psychological perspective. This theory is applicable to the findings of this paper as well. Among the topics with a high proportion of female authors, health information/communities indicate a concern for people as individuals and a focus on interpersonal relationships, and school libraries/information literacy is pertinent to people's education. Research has also demonstrated that males are more engaged in addressing scientific problems while women are more concerned with social issues (Zhang et al., 2021a), which makes it understandable why there are so many men working on the topics of information retrieval/models and algorithms and science evaluation (Chu, 2015).

The primary finding highlights notable gender disparities in research method usage, even within the same research topics. For instance, female authors prefer to use interview within five topics, while male authors tend to use theoretical approach within four topics. As the research methods in this paper are categorized according to data collection techniques, these differences cannot be explained solely by a preference for qualitative or quantitative methods (Ashmos Plowman & Smith, 2011). Social epistemological theories provide insight into these discrepancies. Ashmos Plowman & Smith (2011) suggested that differences in the way men and women perceive may impact their research method selection. Belenky et al.'s (1986) early feminist research revealed that men tend to adopt independent cognitive styles, while women tend to exhibit associative cognitive styles. In associative cognition, individuals acquire knowledge by interacting closely with sources, actively listening and comprehending perspectives, and placing trust in the information received. In contrast, independent cognition involves studying authoritative sources, applying reasoning and analysis to the subject matter, and making independent judgments and conclusions. Interview and theoretical approach align with these cognitive styles, as the former involves interpersonal interaction to gather information, while the latter involves analytical reasoning and theory-based examination.

Furthermore, individuals' early educational experiences exert a significant impact on their methodological preferences. The utilization of certain methods necessitates a considerable level of mathematical confidence, potentially serving as a hypothesized factor influencing individuals' choices. Unfortunately, the prevailing stereotype that females possess lesser aptitude for mathematical learning contributes to a gender disparity in early educational choices (Ceci et al., 2014; Ceci & Williams, 2011). Consequently, even in cases where there is no disparity in mathematical abilities between genders, women exhibit a lower propensity than men to pursue further education in mathematically intensive fields (Zawistowska & Sadowski, 2019). Although current theories and hypotheses offer partial explanations for gender differences, a comprehensive examination of the issue from various perspectives is necessary in the future.

### *6.2 Implications*

Surveys conducted across different journals and topics reveal significant disparities in the utilization of certain research methods between male and female scholars. The findings suggest that female scholars have a wider range of significantly preferred research methods than male scholars, such as interview, questionnaire, and content analysis. However, it is essential for researchers to assess the suitability of a research method for a specific topic instead of relying on personal biases. Having a preference for a particular research method or methods can potentially result in overlooking more innovative or applicable methods during the research process. Giving attention to this issue will positively influence both the study's execution and its subsequent review.

Scholars tend to select research methods in which they are proficient (Nunkoo et al., 2020; Williams et al., 2018), indicating possible disparities between female and male scholars at the time of mastering the methods. This underscores the need for encouraging students to learn research methods extensively and to challenge gender-based learning prejudices and anxieties during teaching. However, as indicated in the discussion, the difference in interest or cognitive style, which cannot be completely avoided, may also be the cause of why men and women have distinct preferences in the choice of research topics and methods. To narrow the gender gap in scientific research output, cooperation can be a viable solution, as mixed research methods are becoming increasingly popular (Järvelin & Vakkari, 2021; Lund & Wang, 2021) and cross-gender collaboration has been demonstrated to promote higher innovation and influence (Yang et al., 2022). Research indicates that women excel at building relationships and being collaborative compared to men (Fell & König, 2016; Moore & Mamiseishvili, 2012). Leveraging the strength allows them to bridge the gap with men, both in terms of the quantity of scientific outputs and the overall scientific impact they achieve.

### *6.3 Limitations*

Although the three selected journals provide some insight into gender differences in topic and method selection within the field of LIS, their representativeness is limited, and it is crucial to include a more extensive range of journals in future research for better generalizability. However, this study primarily utilized a quantitative approach to visualize gender differences, but it overlooked potential gender biases inherent in this method or within the dataset itself. Introducing a qualitative approach, such as interviews, could help address this limitation and provide a more comprehensive understanding of the nuances surrounding gender differences. Additionally, the method of assigning gender based solely on names results in a binary gender label, disregarding the full spectrum of gender identities.

## 7. Conclusion and future work

This study highlights the significant differences between female and male scholars in the use of research methods, even in different topics or journals. While previous research has suggested a general preference for qualitative methods among women and quantitative methods among men, a more nuanced picture where women tend to prefer interviews, while men tend to prefer theoretical approaches was revealed. This discovery opens up new avenues for investigating the underlying causes of these differences in the context of specific research design processes.

Moreover, this study presents the CogFT, a classification model that automatically identifies research methods from full-text data. The accuracy and efficiency of CogFT offers new possibilities for large-scale analyses of research methods, without the need for extensive manual coding. Overall, this study contributes to the research on gender and research methodology, with implications for researchers, academic institutions, and policy-makers seeking to promote gender equity and diversity in academic research.

In future research, it is essential to expand the scope of the study by including a greater number of journals. Doing so will provide a more comprehensive understanding of the gender disparities in topic and method selection across the top journals in the LIS field. To establish a solid foundation, representative studies like Järvelin & Vakkari (2021) can serve as valuable references. Additionally, conducting interview studies with both female and male authors will offer deeper insights into the underlying reasons for these gender differences. Finally, building upon the initial findings of gender disparities across various journals, it would be beneficial to explore the impact of editorial board composition and reviewer selection on the gender composition of contributors in each journal. This consideration could contribute to addressing and rectifying any biases present in the publishing process.

## Acknowledgments

This work was supported by the National Natural Science Foundation of China (Grant No.72074113) and Key Project of Jiangsu Provincial Social Science Foundation (Grant No. 20TQA001). Thanks are due to Professor Heting Chu for providing the annotated data of research methodology for this study.

## Appendix A. Details in identification of research topic.

### A.1 CV of different number of topics.

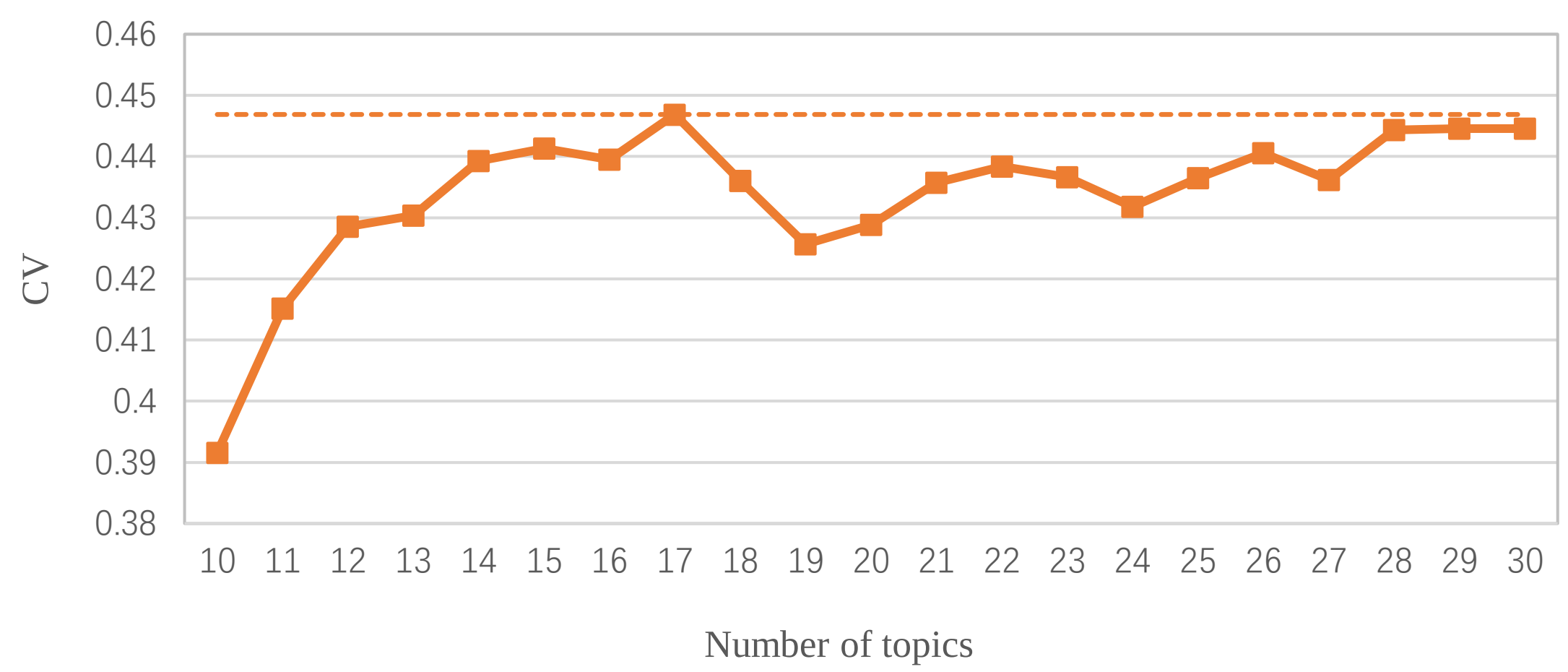

### A.2 Research topics and the top 10 descriptive keywords.

| No. | Research topic | Keywords (Top10) |
|---|---|---|
| T1 | Information seeking behavior | searcher, search, interface, task, query, user, reformulate, browsing, browse, session |
| T2 | Information retrieval/Models and algorithms | vector, parameter, weight, computation, Rocchio (Rocchio algorithm), nonrelevant, probabilistic, optimal, algorithm, performs |
| T3 | Information ethics/Philosophy of information | reality, understood, notion, meaning, epistemology, discursive, argues, philosophical, tradition, epistemological |
| T4 | Science evaluation | fractional, Egghe (Leo Egghe), citation, Lotka, lotkaian, citedness, Rousseau (Ronald Rousseau), all-sciences, Burrell (Quentin L Burrell), publication |
| T5 | Text processing | corpus, word, lexical, Part-of-speech, lexicon, OOV (Handling out of vocabulary), morphological, grammatical, dictionary, noun |
| T6 | School libraries/Information literacy | student, teacher, skill, think, classroom, Kuhlthau (Carol Kuhlthau), taught, instruction, learn, instructional |
| T7 | Bibliometrics/Scientometrics | co-citation, Leydesdorff (Loet Leydesdorff), Boyack (Kevin Boyack), specialty, Blondel (Vincent Blondel), scientometrics, citation, Pajek, Rafols (Ismael Rafols) bibliometric |
| T8 | Health information/Communities | people, health, illness, mother, care, doctor, woman, every day, patient, life |
| T9 | Information services/ Information behavior | credibility, credible, Fogg (Fogg"s Behavior Model), chat, verifiability, socioemotional, question, librarian, answer, questioner |
| T10 | Government policy/statutes | legislation, government, federal rights, society, statutory, denial, statutes, establishment, policy |
| T11 | Organizational information activities/Knowledge management | organizational, employee, manager, consultant, enterprise, organization, business, leadership, management, strategic |
| T12 | Open access publishing | publisher, subscription, publishing, fee, electronic, oa, apc (Article processing charge), Harnad (Stevan Harnad), pricing, charge |
| T13 | Information organization/ | metadata, RDF, XML-based, XML, specification, |
| | Information management | Dublin, extensible, FRBR (Functional Requirements for Bibliographic Records), MARC (Machine Readable Catalog), SPARQL |
| T14 | Technology adoption/ Behavior issues | attitudinal, continuance, Venkatesh (Viswanath Venkatesh), confirmatory, Ajzen (Lcek Ajzen), perceive, tam, intention, PLS (PLS-regression), PBC (perceived behavioral control) |
| T15 | Science of LIS | LIS, JASIS(jasist), discipline, Cronin (Blaise Cronin), librarianship, publish, science, quarterly, faculty, SLIS (School of Library and Information Science) |
| T16 | Personal digital curation/ Personal information management | digital, curation, archival, curating, material, archivist, archive, PIM (Personal information management) , preservation, digitally |
| T17 | Webometrics | webometric, Kousha (Kayvan Kousha), scopus, Thelwall (Mike Thelwall), count, citation, inlinks, Mendeley, webometrics, WIF (Web Impact Factor) |

## Appendix B. Details in classification of research method.

### B.1 Definitions of 16 research methods in Chu & Ke (2017).

| Research method | Definitions |
|---|---|
| Bibliometrics | Bibliometrics is a method of collecting data on publications and citations. It includes citation analysis, informetrics, and scientometrics, depending on the research focus. |
| Content analysis | Content analysis is a method of gathering data by systematically examining texts or other context-based passages. |
| Delphi study | Delphi study is a method of collecting data by seeking advice from a group of experts on a problem to be solved and predicted, reaching consensus through multiple rounds of feedback. |
| Ethnography/field study | Ethnography and field study are data collection methods that involve immersing oneself in the natural settings where participants live or work. Techniques such as observation and interview are applied in this process. |
| Experiment | Experiment is a method of collecting data through systematic procedures that test studied elements in either a laboratory or field setting. In LIS, it is primarily conducted to evaluate new procedures, algorithms, or systems reported in publications. |
| Focus groups | Focus group refers to a method of collecting data through discussion of a research question between a moderator and a group of participants, with emphasis placed on the discussion and interaction among group members. |
| Historical method | Historical method refers to the collection of data by examining, synthesizing, summarizing, and interpreting existing published and unpublished materials related to historical issues. |
| Interview | Interviews is a method of data collection where individual participants are asked questions related to a research problem. The purpose is to gather information about their perspectives, experiences, opinions, and knowledge. |
| Observation | Observation is a method of collecting data by carefully and attentively observing and recording the subject of study. |
| Questionnaire | Questionnaire is a method of collecting data using a list of predefined questions related to the objectives of the study. |
| Research diary/journal | Research diary/journal is a method used by a person who keeps a diary over a period of time to collect data about events, activities, thoughts, reflections or other aspects. |
| Theoretical approach | Theoretical approach is a method of collecting data through conceptual analysis, theoretical examination or similar activities such as conceptual analysis, modelling and theory building. |
| Think aloud protocol | Thinking aloud protocol is a method for collecting data on cognitive activities of participants as they participate in an experiment or perform some task to verbally report their thoughts. |
| Transaction log analysis | Transaction log analysis refers to researchers collecting data by analyzing transaction logs that are automatically captured at either the server or client side. |
| Webometrics | Webometrics is defined as bibliometrics in the web environment, where webpages and websites are generally regarded as publications; with inlinks being considered as citations and outlinks being considered as references. |
| Other methods | Research methods in addition to the 15 methods mentioned above. |

**B.2 Performance of classification models of research method.**

| Model | Macro-P | Macro-R | Macro-F1 | Micro-P | Micro-R | Micro-$F_1$ | Samples-P | Samples-R | Samples-$F_1$ |
|---|---|---|---|---|---|---|---|---|---|
| MS | 0.5340 | 0.4549 | 0.4739 | 0.5791 | 0.4448 | 0.5031 | 0.5791 | 0.5145 | 0.5335 |
| Ngram+Chi2 +NB (CC) | 0.4420 | 0.7406 | 0.5431 | 0.4150 | 0.7486 | 0.5340 | 0.5211 | 0.7704 | 0.5794 |
| SciBERT-Truncation | 0.8526 | 0.7698 | 0.8031 | 0.8190 | 0.7624 | 0.7897 | 0.8375 | 0.8077 | 0.8079 |
| SciBERT-Hierarchical | 0.8082 | 0.7487 | 0.7590 | 0.7637 | 0.7680 | 0.7658 | 0.7986 | 0.8159 | 0.7888 |
| BigBird | **0.8325** | 0.7902 | 0.8043 | 0.8270 | 0.7790 | 0.8023 | 0.8527 | 0.8261 | 0.8251 |
| CogFT | 0.8266 | **0.8434** | **0.8299** | **0.8137** | **0.8204** | **0.8171** | **0.8756** | **0.8748** | **0.8551** |

**B.3 Results of the research method classification using CogFT.**

| No. | Research method | Frequency | Percent |
|---|---|---|---|
| 1 | Experiment | 1,314 | 19.04 |
| 2 | Theoretical approach | 989 | 14.33 |
| 3 | Questionnaire | 912 | 13.21 |
| 4 | Content analysis | 899 | 13.02 |
| 5 | Bibliometrics | 870 | 12.61 |
| 6 | Interview | 690 | 10.00 |
| 7 | Transaction log analysis | 260 | 3.77 |
| 8 | Observation | 218 | 3.16 |
| 9 | Webometrics | 143 | 2.07 |
| 10 | Other methods | 129 | 1.87 |
| 11 | Historical method | 122 | 1.77 |
| 12 | Think aloud protocol | 113 | 1.64 |
| 13 | Focus groups | 89 | 1.29 |
| 14 | Ethnography/field study | 88 | 1.27 |
| 15 | Research diary/journal | 50 | 0.72 |
| 16 | Delphi study | 16 | 0.23 |
| | Total | 6902 | 100 |

**Appendix C. Statistical results of gender differences in topic and method selection.**

**C.1 Comparison of the share of gender-specific authors in different topics.**

| Research topic | Share of papers in the topic (%) | Share of female authored papers （%） | Share of male authored papers （%） | Gender×Topic Chi-square |
|---|---|---|---|---|
| Information seeking behavior | 10.13 | **51.31** | 48.69 | 32.1094*** |
| Information retrieval/Models and algorithms | 9.43 | 18.11 | **81.89** | 93.8199*** |
| Information ethics/Philosophy of information | 8.12 | 32.48 | **67.52** | 8.4305** |
| Science evaluation | 7.07 | 15.82 | **84.18** | 86.4335*** |
| Text processing | 6.66 | 32.19 | **67.81** | 7.4962** |
| School libraries/Information literacy | 6.56 | **58.38** | 41.62 | 52.6157*** |
| Bibliometrics/Scientometrics | 6.09 | 26.17 | **73.83** | 23.3125*** |
| Health information/Communities | 5.82 | **71.66** | 28.34 | 134.4652*** |
| Information services/ Information behavior | 5.40 | **47.02** | 52.98 | 7.0540** |
| Government policy/statutes | 5.04 | 38.72 | 61.28 | 0.0416 |
| Organizational information activities/KM | 4.97 | **46.56** | 53.44 | 5.7433* |
| Open access publishing | 4.89 | 34.50 | 65.5 | 2.5290 |
| Information organization/IM | 4.63 | 32.79 | **67.21** | 4.3810* |
| Technology adoption/ Behavior issues | 4.17 | **47.27** | 52.73 | 5.8128* |
| Science of LIS | 4.06 | 45.79 | 54.21 | 3.7448 |
| Personal digital curation/Personal IM | 3.51 | **59.46** | 40.54 | 31.4067*** |
| Webometrics | 3.45 | 29.12 | **70.88** | 7.9533** |
| Total | 100 | **39.33** | **60.67** | |

**C.2 Comparison of the share of gender-specific authors in articles using different methods.**

| Research method | Share of papers containing the method(%) | Share of female authored papers （%） | Share of male authored papers （%） | Gender×Method Chi-square |
|---|---|---|---|---|
| Experiment | 24.86 | 33.94 | **66.06** | 15.9549*** |
| Theoretical approach | 18.74 | 30.16 | **69.84** | 34.8206*** |
| Questionnaire | 17.26 | **54.18** | 45.82 | 84.0232*** |
| Content analysis | 16.99 | **45.54** | 54.46 | 14.4491*** |
| Bibliometrics | 16.50 | 25.17 | **74.83** | 73.1039*** |
| Interview | 13.07 | **65.60** | 34.40 | 199.2635*** |
| Transaction log analysis | 4.93 | 44.23 | 55.77 | 2.6143 |
| Observation | 4.13 | **64.22** | 35.78 | 56.5874*** |
| Webometrics | 2.71 | 32.87 | 67.13 | 2.5052 |
| Other methods | 2.45 | 44.19 | 55.81 | 1.2735 |
| Historical method | 2.31 | 36.07 | 63.93 | 0.5457 |
| Think aloud protocol | 2.14 | **65.49** | 34.51 | 32.3934*** |
| Focus groups | 1.67 | **57.95** | 42.05 | 12.7889*** |
| Ethnography/field study | 1.67 | **61.36** | 84.18 | 17.8999*** |
| Research diary/journal | 0.95 | **70.00** | 30.00 | 19.7070*** |
| Delphi study | 0.30 | 37.50 | 62.50 | 0.1725 |
| Total | ^ | **39.33** | **60.67** | |

**Note**: ^: It is not 100. This is because the "share of papers containing the method" is equal to the number of articles that employed the method divided by the total number of articles (5273) rather than the total frequency of method use. Detailed explanation can be found in the fourth paragraph of section 5.1.

## C.3 A two-dimensional heat map based on the PFA in topics and methods.

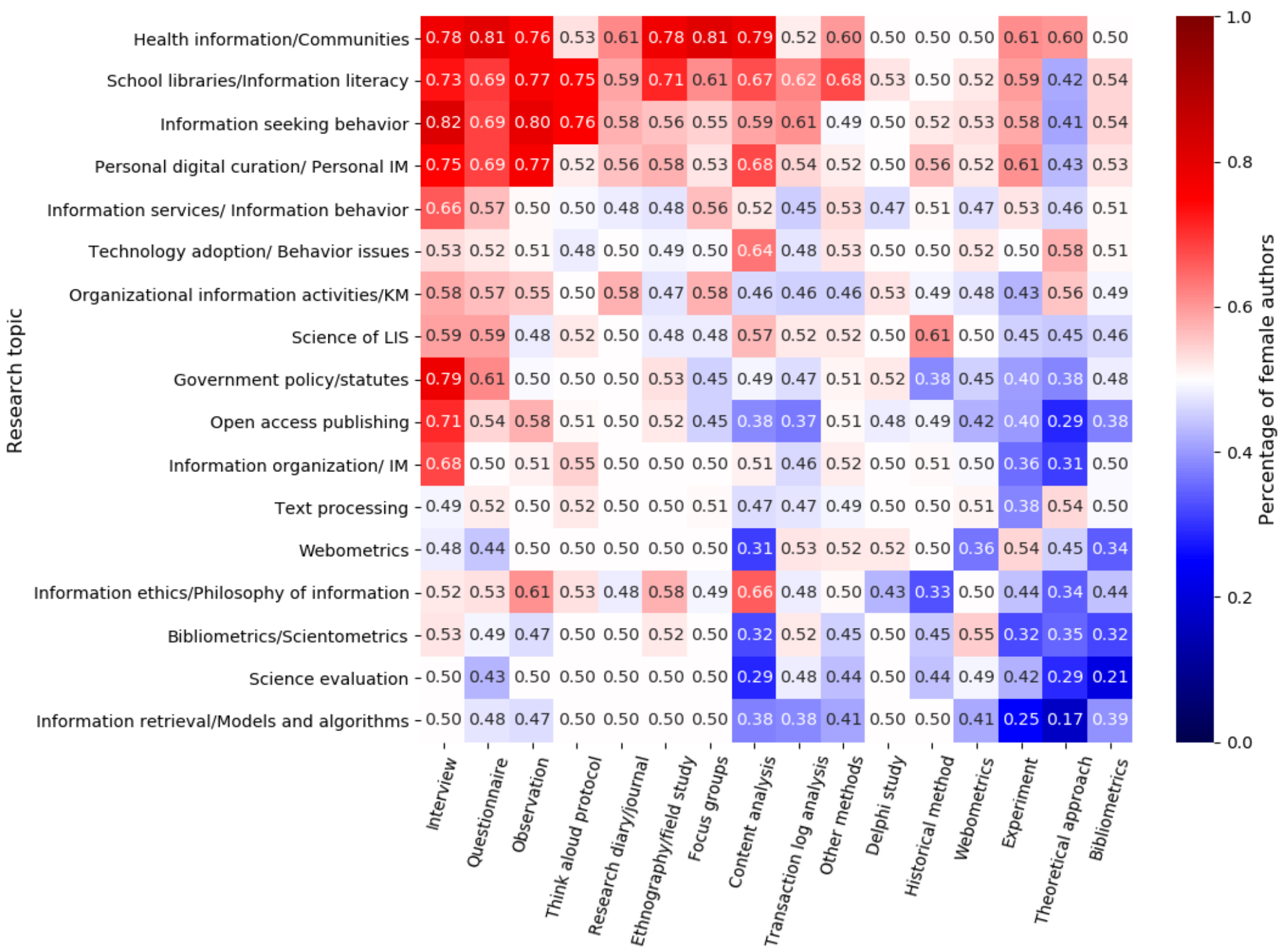

## C.4 Distribution of gender differences in method usage within research topics.

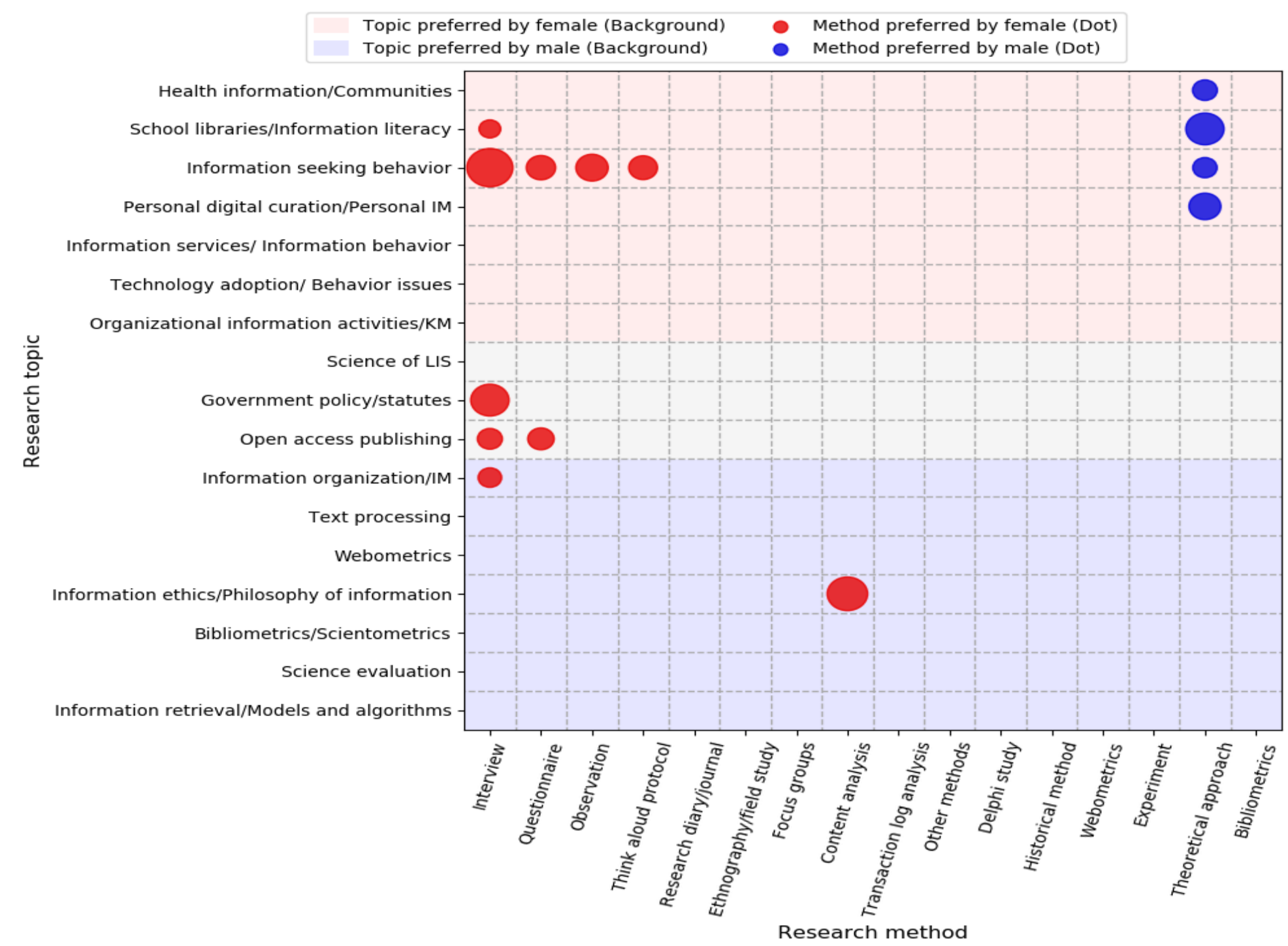

## C.5 Gender and topic-based frequency distribution of JASIST, JDoc, and LISR papers.

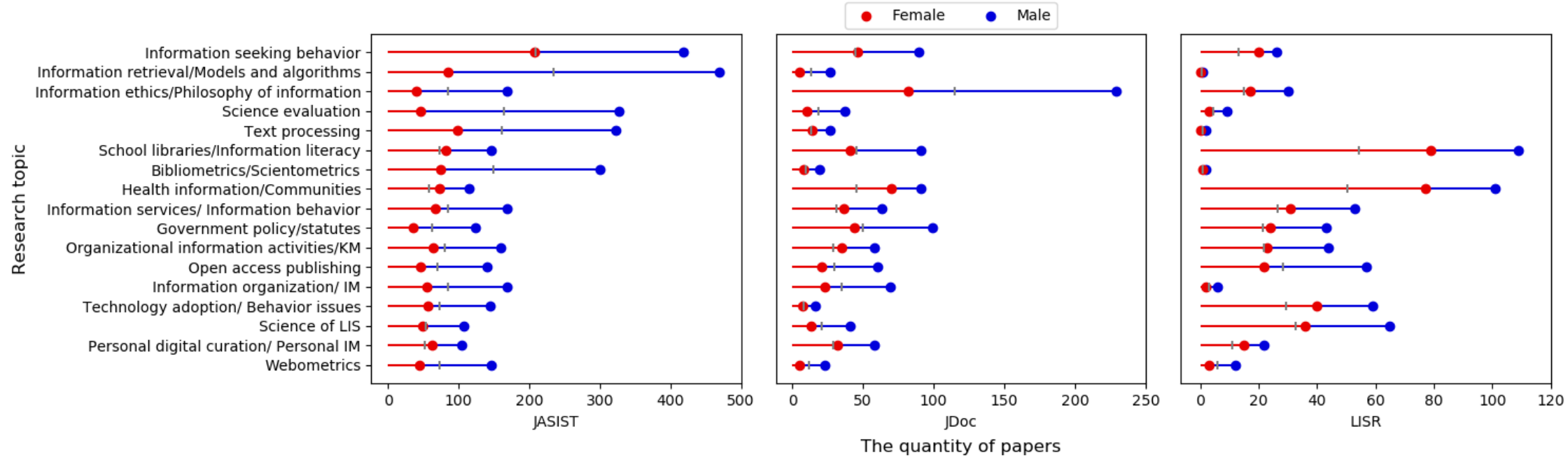

## C.6 Gender and method-based frequency distribution of JASIST, JDoc, and LISR papers.

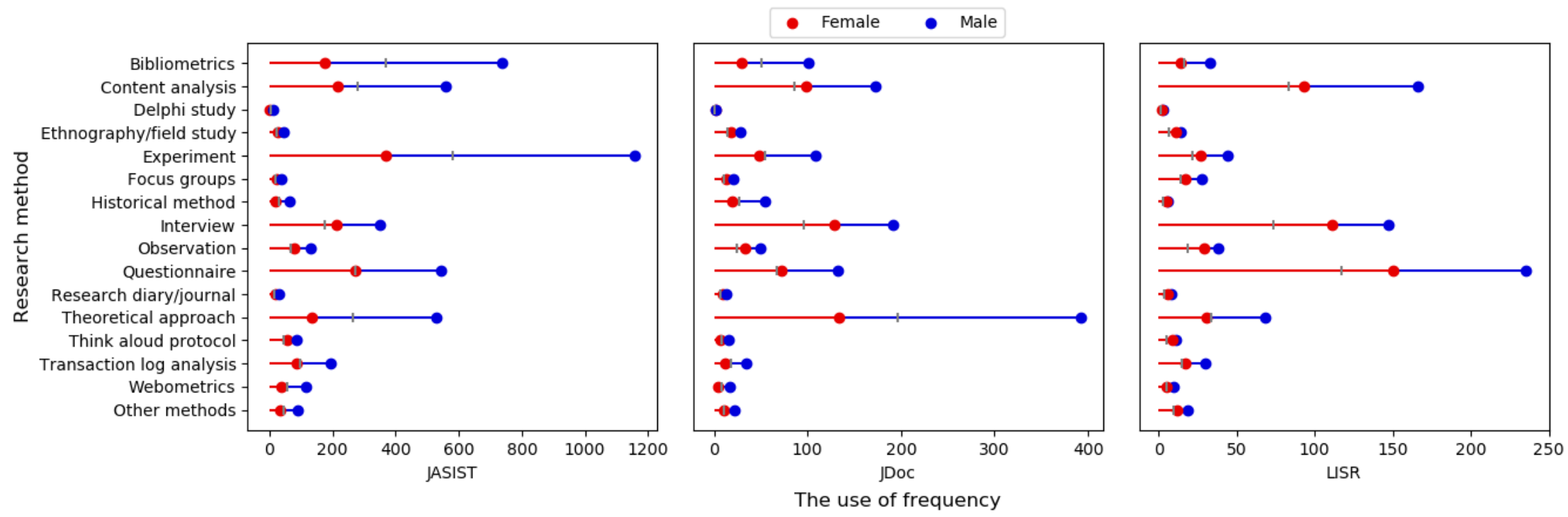